RESEARCH ARTICLE

# Continuous Distributed Representation of Biological Sequences for Deep Proteomics and Genomics

**Ehsaneddin Asgari[1], Mohammad R. K. Mofrad[1,2]***

**1** Molecular Cell Biomechanics Laboratory, Departments of Bioengineering and Mechanical Engineering, University of California, Berkeley, California 94720, United States of America, **2** Physical Biosciences Division, Lawrence Berkeley National Lab, Berkeley, California 94720, United States of America

* mofrad@berkeley.edu

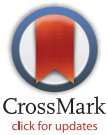

**Data Availability Statement:** All relevant data are within the paper and its Supporting Information files. Our web-based tools and trained data are available at Life Language Processing website: http://llp.berkeley. edu and through the Dataverse database: http://dx. doi.org/10.7910/DVN/JMFHTN, and will be regularly updated for calculation/classification of ProtVecs as well as visualization of biological sequences.

**Funding:** Financial support from National Science Foundation through a CAREER Award (CBET-0955291) is gratefully acknowledged. The funders had no role in study design, data collection and analysis, decision to publish, or preparation of the manuscript.

## Abstract

We introduce a new representation and feature extraction method for biological sequences. Named bio-vectors (BioVec) to refer to biological sequences in general with protein-vectors (ProtVec) for proteins (amino-acid sequences) and gene-vectors (GeneVec) for gene sequences, this representation can be widely used in applications of deep learning in proteomics and genomics. In the present paper, we focus on protein-vectors that can be utilized in a wide array of bioinformatics investigations such as family classification, protein visualization, structure prediction, disordered protein identification, and protein-protein interaction prediction. In this method, we adopt artificial neural network approaches and represent a protein sequence with a single dense n-dimensional vector. To evaluate this method, we apply it in classification of 324,018 protein sequences obtained from Swiss-Prot belonging to 7,027 protein families, where an average family classification accuracy of 93% ± 0.06% is obtained, outperforming existing family classification methods. In addition, we use ProtVec representation to predict disordered proteins from structured proteins. Two databases of disordered sequences are used: the DisProt database as well as a database featuring the disordered regions of nucleoporins rich with phenylalanine-glycine repeats (FG-Nups). Using support vector machine classifiers, FG-Nup sequences are distinguished from structured protein sequences found in Protein Data Bank (PDB) with a 99.8% accuracy, and unstructured DisProt sequences are differentiated from structured DisProt sequences with 100.0% accuracy. These results indicate that by only providing sequence data for various proteins into this model, accurate information about protein structure can be determined. Importantly, this model needs to be trained only once and can then be applied to extract a comprehensive set of information regarding proteins of interest. Moreover, this representation can be considered as pre-training for various applications of deep learning in bioinformatics. The related data is available at Life Language Processing Website: http://llp. berkeley.edu and Harvard Dataverse: http://dx.doi.org/10.7910/DVN/JMFHTN.

## Introduction

Nature uses certain languages to describe biological sequences such as DNA, RNA, and proteins. Much like humans adopt languages to communicate, biological organisms use sophisticated languages to convey information within and between cells. Inspired by this conceptual analogy, we adopt existing methods in natural language processing (NLP) to gain a deeper understanding of the "language of life" with the ultimate goal to discover functions encoded within biological sequences [1–4].

Feature extraction is an important step in data analysis, machine learning and NLP. It refers to finding an interpretable representation of data for machines that can increase performance of learning algorithms. Even the most sophisticated algorithms would perform poorly if inappropriate features are used, while simple methods can potentially perform well when they are fed with the appropriate features. Feature extraction can be done manually or in an unsupervised fashion. In this paper, we propose an unsupervised data-driven distributed representation for biological sequences. This method, called bio-vectors (BioVec) in general and more specifically protein-vectors (ProtVec) for proteins, can be applied to a wide range of problems in bioinformatics, such as protein visualization, protein family classification, structure prediction, domain extraction, and interaction prediction. In this approach, each biological sequence is embedded in an n-dimensional vector that characterizes biophysical and biochemical properties of sequences using neural networks. In the following, we first explain how this method works and how it is trained from 546,790 sequences of Swiss-Prot database. Subsequently, we will analyze the biophysical and the biochemical properties of this representation qualitatively and quantitatively. To further evaluate this feature extraction method, we apply it in classification of 7,027 protein families of 324,018 protein sequences in Swiss-Prot. In the next step, we use this approach for visualization and characterization of two categories of disordered sequences: the DisProt database as well as a database of disordered regions of phenylalanine-glycine nucleoporins (FG-Nups). Finally, we classify these protein families using support vector machine (SVM) classifiers [5]. As a key advantage of the proposed method, the embedding needs to be trained only once and then may be used to encode biological sequences in a given problem. The related data and future updated will be available at: http://llp.berkeley.edu and Harvard Dataverse: http://dx.doi.org/10.7910/DVN/JMFHTN.

## Distributed Representation

Distributed representation has proved one of the most successful approaches in machine learning [6–10]. The main idea in this approach is encoding and storing information about an item within a system through establishing its interactions with other members. Distributed representation was originally inspired by the structure of human memory, where the items are stored in a "content-addressable" fashion. Content-based storing allows for efficiently recalling items from partial descriptions. Since the content-addressable items and their properties are stored within a close proximity, such a system provides a viable infrastructure to generalize features attributed to an item.

Continuous vector representation, as a distributed representation for words, has been recently established in natural language processing (NLP) as an efficient way to represent semantic/syntactic units with many applications. In this model, each word is embedded in a vector in an n-dimensional space. Similar words have close vectors, where similarity is defined in terms of both syntax and semantic. The basic idea behind training such vectors is that the meaning of a word is characterized by its context, i.e. neighboring words. Thus, words and their contexts are considered to be positive training samples. Such vectors can be trained using large amounts of textual data in a variety of ways, e.g. neural network architectures like the Skip-gram model [10].

Interesting patterns have been observed by training word vectors using Skip-gram in natural language. Words with similar vector representations show multiple degrees of similarity. For instance, $\overrightarrow{King} - \overrightarrow{Man} + \overrightarrow{Woman}$ resembles the closest vector to the word $\overrightarrow{Queen}$ [11].

In this work, we seek unique patterns in biological sequences to facilitate biophysical and biochemical interpretations. We show how Skip-gram can be used to train a distributed representation for biological sequences over a large set of sequences, and establish physical and chemical interpretations for such representations. We propose this as a general-purpose representation for protein sequences that can be used in a wide range of bioinformatics problems, including protein family classification, protein interaction prediction, structure prediction, motif extraction, protein visualization, and domain identification. To illustrate, we specifically tackle visualization and protein family classification problems.

## Protein Family Classification

A protein family is a set of proteins that are evolutionarily related, typically involving similar structures or functions. The large gap between the number of known sequences versus the amount of known functional information about sequences has motivated family (function) identification methods based on primary sequences [12–14]. Protein Family Database (Pfam) is a widely used source for protein families [15]. In Pfam, a family can be classified as a "family", "domain", "repeat", or "motif". In this study, we utilize ProtVec to classify protein families in Swiss-Prot using the information provided by Pfam database and we obtain a high classification accuracy.

Protein family classification based on the primary structures (sequences) has been performed using classifiers such as support vector machine classifier (SVM) [16–18]. Besides the primary sequence, the existing methods typically require extensive feature extraction, e.g. hydrophobicity, normalized Van der Waals volume, polarity, polarizability, charge, surface tension, secondary structure and solvent accessibility. The reported accuracies of a previous study on family classification have been in the range of 69.1–99.6% for 54 protein families [16]. In another study, researchers used motifs from protein interactions for detecting Structural Classification of Proteins (SCOP) [19] families for 368 proteins, and obtained a classification accuracy of 75% at sensitivity of 10% [20]. In contrast, our proposed approach is trained based solely on primary sequence information, yet achieving high accuracy when applied in classifications of protein families.

## Disordered Proteins

Proteins can be fully or partially unstructured, i.e. lacking a secondary or ordered tertiary three-dimensional structure. Due to their abundance and the critical roles they play in cell biology, disordered proteins are considered to be an important class of proteins [21]. Several studies have focused on disordered peptides and their functional analysis in recent years [22–24].

In the present work, we introduce ProtVec for the visualization and characterization of two categories of disordered proteins: DisProt database as well as a database of disordered regions of phenylalanine-glycine nucleoporins (FG-Nups) [25].

DisProt is a database of experimentally identified disordered proteins that categorizes disordered and ordered regions of a collection of proteins [26]. DisProt Release 6.02 consists of 694 proteins presenting 1539 disordered, and 95 ordered regions. FG-Nups dataset is a collection of FG-Nups disordered sequences [27]. Nucleoporins form the nuclear pore complex (NPC), the sole gateway for bidirectional transport of cargo between the cytoplasm and the nucleus in eukaryotic cells [28]. Since FG-Nups are mostly computationally identified, only 10 sequences out of 1,138 disordered sequences exist in Swiss-Prot. A recent study on features of FG-Nups

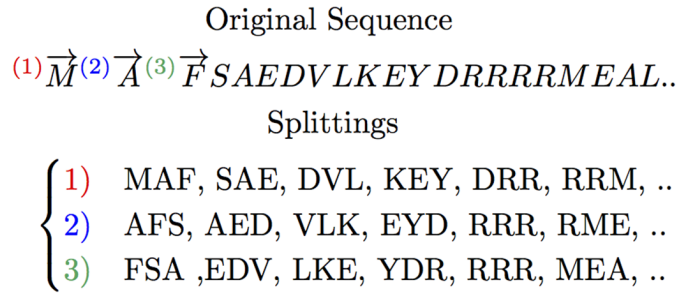

Fig 1. Protein sequence splitting. In order to prepare the training data, each protein sequence will be represented as three sequences (1, 2, 3) of 3-grams.

doi:10.1371/journal.pone.0141287.g001

versus DisProt showed biophysical differences between FG-Nups and average DisProt sequences [29].

We further propose using protein-vectors for the visualization of biological sequences. Simplicity and biophysical interpretations encoded within ProtVec distinguishes this method from the previous work [30, 31]. As an example, we use ProtVec for the visualization of FG-Nups, DisProt, and structured PDB proteins. This visualization confirms the results obtained [29] on the biophysical features of FG-Nups and typical disordered proteins. Furthermore, we employ ProtVec to classify FG-Nups versus random PDB sequences as well as DisProt disordered regions versus disport ordered regions.

## Methods

### Protein-Space Construction

Our goal is to construct a distributed representation of biological sequences. In the training process of word embedding in NLP, a large corpus of sentences should be fed into the training algorithm to ensure sufficient contexts are observed. Similarly, a large corpus is needed to train distributed representation of biological sequences. We use Swiss-Prot as a rich protein database, which consists of 546,790 manually annotated and reviewed sequences.

The next step in training distributed representations is to break the sequences into sub sequences (i.e. biological words). The simplest and most common technique in bioinformatics to study sequences involves fixed-length overlapping n-grams [32–34]. However, instead of using n-grams directly in feature extraction, we utilize n-gram modeling for training a general purpose distributed representation of sequences. This so-called embedding model needs to be trained only once and may then be adopted in feature extraction part of specific problems.

In n-gram modeling of protein informatics, usually an overlapping window of 3 to 6 residues is used. Instead of taking overlapping windows, we generate 3 lists of shifted non-overlapping words, as shown in Fig 1. Evaluating K-nearest neighbors in a 2xfold cross-validation for different window sizes, embedding vector sizes and overlapping versus non-overlapping n-grams showed a more consistent embedding training for a window size of 3 and the mentioned splitting.

The same procedure is applied on all 546,790 sequences in Swiss-Prot, thus at the end we obtain a corpus consisting of $546,790 \times 3 = 1,640,370$ sequences of 3-grams (3-gram is a "biological" word consisting of 3 amino acids). The next step is training the embedding based on such data through a Skip-gram neural network. In training word vector representations, Skip-gram attempts to maximize the probability of observed word sequences (contexts). In other words, for a given training sequence of words we would like to find their corresponding n-dimensional vectors maximizing the following average log probability function. Such a constraint allows similar words to assume a similar representation in this space.

$$\underset{v, v'}{\arg\max} \frac{1}{N} \sum_{i=1}^{N} \sum_{-c \leq j \leq c, j \neq 0} \log p(w_{i+j} | w_i)$$

$$p(w_{i+j} | w_i) = \frac{\exp(v_{w_{i+j}}^{'T} v_{w_i})}{\sum_{k=1}^{W} \exp(v_{w_k}^{'T} v_{w_i})},$$

(1)

where $N$ is the length of the training sequence, $2c$ is the window size we consider as the context, $w_i$ is the center of the window, $W$ is the number of words in the dictionary and $v_w$ and $v'_w$ are input and output n-dimensional representations of word $w$, respectively. The probability $p(w_{i+j} | w_i)$ is defined using a softmax function. Hierarchical softmax or negative sampling are efficient approximations of such a softmax function. In the implementation we use (Word2Vec) [10] negative sampling has been utilized, which is considered as the state-of-the-art for training word vector representation. Negative sampling uses the following objective function in the calculation of the word vectors:

$$\underset{\theta}{\arg\max} \prod_{(w,c) \in D} p(D = 1 | c, w; \theta) \prod_{(w,c) \in D'} p(D = 0 | c, w; \theta),$$

(2)

where $D$ is the set of all word and context pairs $(w, c)$ existing in the training data (positive samples) and $D'$ is a randomly generated set of incorrect $(w, c)$ pairs (negative samples).

$p(D = 1 | w, c; \theta)$ is the probability that $(w, c)$ pair came from the training data and $p(D = 0 | w, c; \theta)$ is the probability that $(w, c)$ did not come from the training data. The term $p(D = 1 | c, w; \theta)$ can be defined using a sigmoid function on the word vectors:

$$p(D = 1 | w, c; \theta) = \frac{1}{1 + e^{-v_c \cdot v_w}},$$

where the parameters $\theta$ are the word vectors we train within the optimization framework: $v_c$ and $v_w \in R^d$ are vector representations for the context $c$ and the word $w$ respectively [35]. In Eq (2), the positive samples maximize the probabilities of the observed $(w, c)$ pairs in the training data, while the negative samples prevent all vectors from having the same value by disallowing some incorrect $(w, c)$ pairs. To train the embedding vectors, we consider a vector size of 100 and a context size of 25. Thus each 3-gram is presented as a vector of size 100.

## Protein-Space Analysis

To qualitatively analyze the distribution of various biophysical and biochemical properties within the training space, we project all 3-gram embeddings from 100-dimensional space to a 2D space using Stochastic Neighbor Embedding [36]. Mass, volume, polarity, hydrophobicity, charge, and van der Waals volume properties were analyzed. The data is adopted from [37]. In addition, to quantitatively measure the continuity of these properties in the protein-space, the best Lipschitz constant, i.e. the smallest $k$ satisfying is calculated:

$$d_f(f_{prop}(w_1), f_{prop}(w_2)) \leq k \times d_w(w_1, w_2),$$

(3)

where $f$ is the scale of one of the properties of a given 3-grams (e.g., average mass, hydrophobicity, etc.), $d$ is the distance metric, $d_f$ is the absolute value of score differences, and $d_w$ is Euclidian distance between two 3-grams $w_1$ and $w_2$. The Lipschitz constant is calculated for the aforementioned properties.

## Protein Family Classification

To evaluate the strength of the proposed representation, we set up a classification task on protein families. Family information of 324,018 protein sequences in Swiss-Prot is extracted from the Protein Family (Pfam) database, resulting in a total of 7,027 distinct families for Swiss-Prot sequences. Each sequence is represented as the summation of the vector representation of overlapping 3-grams. Thus, each sequence is presented as a vector of size 100. For each family type, the same number of instances from Swiss-Prot are selected randomly to form the negative examples. Support vector machine classifiers are used to evaluate the strength of ProtVec in the classification of protein families through 10 × fold cross-validations. We perform the classification over 7,027 protein families consisting of 324,018 sequences. For the evaluation we report specificity (true negative rate), sensitivity (true positive rate), and the accuracy of family classifications.

$$Sensitivity = TP\ rate = \frac{TP}{TP + FN}$$

$$Specificity = 1 - FP\ rate = \frac{TN}{FP + TN}$$

$$Accuracy = \frac{TP + TN}{P + N}$$

## Visualization and Classification of Disordered Proteins

Two databases of disordered proteins are used for disordered protein prediction: DisProt database (694 sequences) and FG-Nups dataset (1,138 sequences).

**FG-Nups Characterization.** To distinguish the characteristics of FG-Nups, a collection of 1,138 FG-Nups and two random sets of 1,138 structured proteins from Protein Data Bank (PDB) [38] are compared. Since PDB sequences on average have a shorter length than disordered proteins, the two sets are selected from PDB in such a way that they have an average length of 900 residues, the same as the average length of the disordered protein dataset. For visualization purposes, the ProtVec is reduced from 100 dimensions to a 2D space using Stochastic Neighbor Embedding [36].

We quantitatively evaluate how ProtVec can be used to distinguish between FG-Nups versus typical PDB sequences using a support vector machine binary classifier. The positive examples were the aforementioned 1,138 disordered FG-Nups proteins and the negative examples (again 1,138 sequences) are selected randomly from PDB with the same average length of disordered sequences ($\approx 900$ residues). We present each protein sequence as a summation of its ProtVecs of all 3-grams. Since the average length of structured proteins is shorter than FG-Nups, and to avoid trivial cases, the PDB sequences are selected in a way to maintain the same average length.

**DisProt Characterization.** To distinguish the characteristics of DisProt sequences, we use DisProt Release 6.02, consisting of 694 proteins presenting 1539 disordered and 95 ordered regions, and perform the same experiment as for FG-Nups with DisProt sequences.

## Results

### Protein-Space Analysis

Although the protein-space is trained based on only the primary sequences of proteins, it offers several interesting biochemical and biophysical implications. In order to study these features,

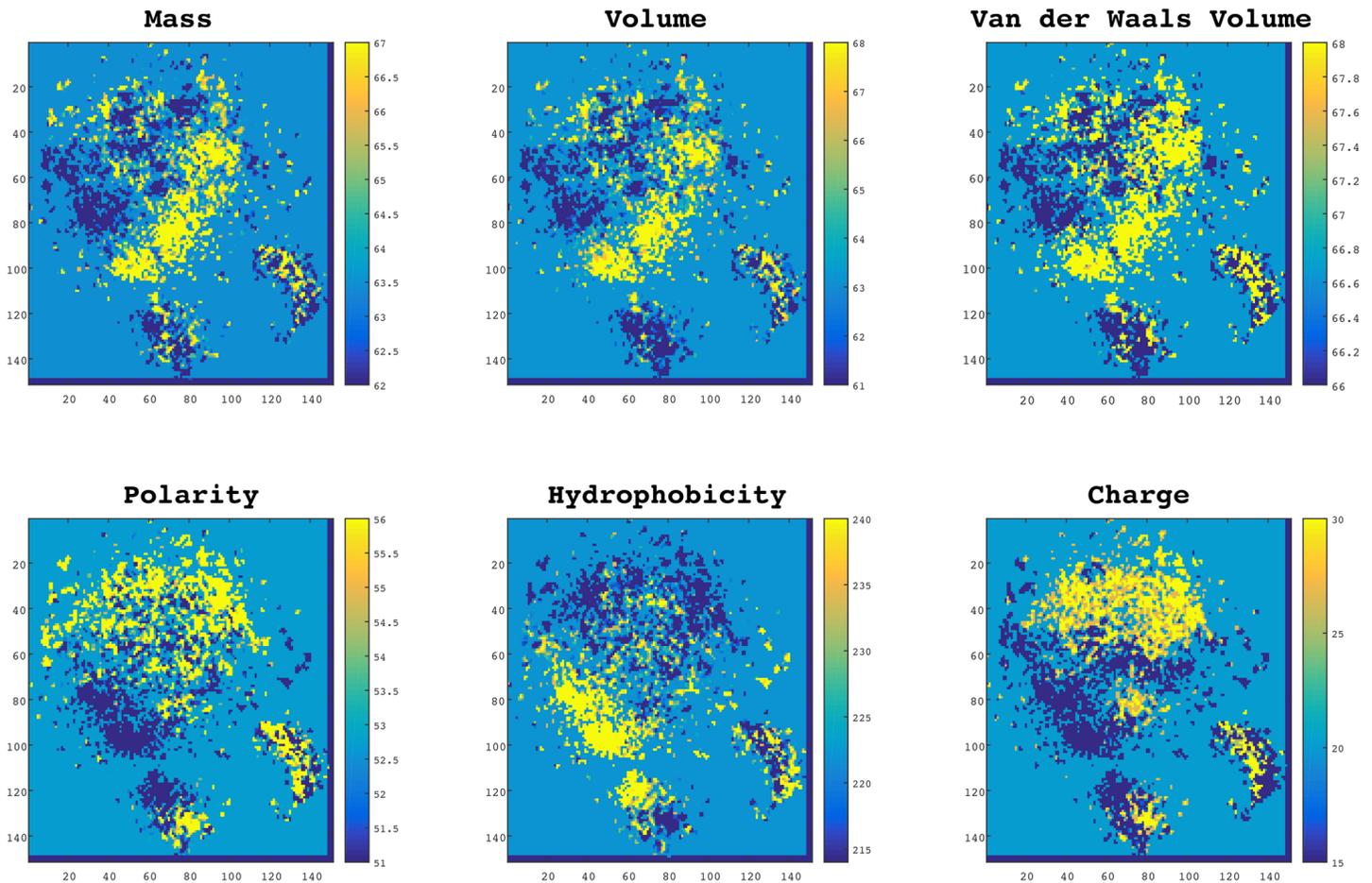

**Fig 2. Normalized distributions of biochemical and biophysical properties in protein-space.** In these plots, each point represents a 3-gram (a word of three residues) and the colors indicate the scale for each property. Data points in these plots are projected from a 100-dimensional space a 2D space using t-SNE. As it is shown words with similar properties are automatically clustered together meaning that the properties are smoothly distributed in this space.

doi:10.1371/journal.pone.0141287.g002

we visualized the distribution of different criteria, including mass, volume, polarity, hydrophobicity, charge, and van der Waals volume in this space. To do so, for each 3-gram we conducted qualitative and quantitative analyses as described below.

**Qualitative Analysis.** In order to visualize the distribution of the aforementioned properties, we projected all 3-gram embeddings from 100-dimensional space to a 2D space using Stochastic Neighbor Embedding (t-SNE) [36]. In the diagrams presented in Fig 2, each point represents a 3-gram and is colored according to its scale in each property. Interestingly, as can be seen in the figure, 3-grams with the same biophysical and biochemical properties were grouped together. This observation suggests that the proposed embedding not only encodes protein sequences in an efficient way that proved useful for classification purposes, but also reveals some important physical and chemical patterns in protein sequences.

**Quantitative Analysis.** Although Fig 2 illustrates the smoothness of protein-space with respect to different physical and chemical meanings, we required a quantitative approach to measure the continuity of these properties in the protein-space. To do so, we calculated the best Lipschitz constant. For all 6 properties presented in Fig 2, we calculated the minimum $k$. To evaluate this result we made an artificial space called "scrambled space" by randomly

**Table 1. Using Lipschitz number to evaluate the continuity of ProtVec with respect to biophysical and biochemical properties.**

| Property | Lipschitz Number | | |
|---|---|---|---|
| | protein-Space | The scrambled space | Ratio |
| Mass | 0.3137 | 0.6605 | 0.4750 |
| Volume | 0.3742 | 0.6699 | 0.5586 |
| Van Der Waal Volume | 0.3629 | 0.6431 | 0.5643 |
| Polarity | 0.4757 | 1.2551 | 0.3790 |
| Hydrophobicity | 0.608 | 1.448 | 0.4203 |
| Charge | 0.8733 | 1.3620 | 0.6412 |
| Average | 0.50 | 1.01 | 0.51 |

doi:10.1371/journal.pone.0141287.t001

shuffling the labels of 3-grams in the 100 dimensional space. Table 1 contains the values of Libschitz constants for protein-space versus the "scrambled space" with respect to different properties and also their ratio.

Normally if $k = 1$ the function is called a short map, and if $0 \leq k < 1$ the function is called a contraction. The results suggest that the protein-space is on average 2-times smoother in terms of physical and chemical properties than a random space. This quantitative result supports our qualitative observation of the space structure in Fig 2, and suggests that our training space encodes, 3-grams in an informative manner.

## Protein Family Classification

In order to evaluate the strength of ProtVec, we performed classifications of 7,027 protein families and obtained a weighted average accuracy of 93 ± 0.06%, which exhibits a more reliable result than the existing methods. In contrast to the existing methods, our proposed approach is trained based on primary sequence information alone.

Table 2 shows the sensitivity, specificity, and the accuracy for the most frequent families in Swiss-Prot. These results suggest that structural features of proteins can be accurately predicted from the primary sequence information solely. The results for all 7,027 families can be found in Supplementary Information, see S1 File. The average accuracy for the first 1,000 (261,149 sequences), 2,000 (293,957 sequences), 3,000 (308,292 sequences), and 4,000 (316,135 sequences) frequent families were respectively 94% ± 0.05%,93% ± 0.05%, 92% ± 0.06%, and 91% ± 0.08%. To compute the overall accuracy for all 7,026 families, we calculated the weighted average accuracy, because for the families with number of instances less than 10, the validation set are not statistically sufficient and they should have less contribution in the overall accuracy. The weighted accuracy of all 7,027 families (weighted based on the number of instances) was 93% ± 0.06%.

## Disordered Proteins Visualization and Classification

Due to the functional importance of disordered proteins, prediction of unstructured regions of disordered proteins and determining the sequence patterns featured in disordered regions is a critical problem in protein bioinformatics. We evaluated the ability of ProtVec to characterize and discern disordered protein sequences from structured sequences.

**FG-Nups Characterization.**   In this case study, we used the FG-Nups collection of 1,138 disordered proteins containing disorder regions with a fraction of at least one third of the sequence length. For comparison purposes, we also collected two sets of structured proteins from Protein Data Bank (PDB).

Table 2. Performance of protein family classification using SVM and ProtVec over some of the most frequent families in Swiss-Prot. Families are sorted with respect to their frequency in Swiss-Prot.

| Family name | Training instances | | Classification Result | | |
|---|---|---|---|---|---|
| | # of positive sequences | # of negative sequences | Specificity | Sensitivity | Accuracy |
| 50S ribosome-binding GTPase | 3,084 | 3,084 | 0.95 | 0.93 | 0.94 |
| Helicase conserved C-terminal domain | 2,518 | 2,518 | 0.83 | 0.80 | 0.82 |
| ATP synthase alpha-beta family, nucleotide-binding domain | 2,387 | 2,387 | 0.98 | 0.97 | 0.97 |
| 7 transmembrane receptor (rhodopsin family) | 1,820 | 1,820 | 0.95 | 0.96 | 0.95 |
| Amino acid kinase family | 1,750 | 1,750 | 0.91 | 0.92 | 0.91 |
| ATPase family associated with various cellular activities (AAA) | 1711 | 1711 | 0.92 | 0.90 | 0.91 |
| tRNA synthetases class I (I, L, M and V) | 1,634 | 1,634 | 0.97 | 0.97 | 0.97 |
| tRNA synthetases class II (D, K and N) | 1,419 | 1,419 | 0.88 | 0.83 | 0.85 |
| Major Facilitator Superfamily | 1,303 | 1,303 | 0.95 | 0.97 | 0.96 |
| Hsp70 protein | 1,272 | 1,272 | 0.97 | 0.97 | 0.97 |
| NADH-Ubiquinone-plastoquinone (complex I), various chains | 1,251 | 1,251 | 0.97 | 0.97 | 0.97 |
| Histidine biosynthesis protein | 1,248 | 1,248 | 0.96 | 0.97 | 0.97 |
| TCP-1-cpn60 chaperonin family | 1,246 | 1,246 | 0.95 | 0.96 | 0.95 |
| EPSP synthase (3-phosphoshikimate 1-carboxyvinyltransferase) | 1,207 | 1,207 | 0.96 | 0.96 | 0.96 |
| Aldehyde dehydrogenase family | 1,200 | 1,200 | 0.93 | 0.94 | 0.94 |
| Shikimate–quinate 5-dehydrogenase | 1,128 | 1,128 | 0.87 | 0.89 | 0.88 |
| GHMP kinases N terminal domain | 1,120 | 1,120 | 0.88 | 0.92 | 0.90 |
| Ribosomal protein S2 | 1,083 | 1,083 | 0.95 | 0.96 | 0.95 |
| Ribosomal protein S4–S9 N-terminal domain | 1,072 | 1,072 | 0.95 | 0.97 | 0.96 |
| Ribosomal protein L16p-L10e | 1,053 | 1,053 | 0.95 | 0.96 | 0.96 |
| KOW motif | 1,047 | 1,047 | 0.93 | 0.95 | 0.94 |
| Uncharacterized protein family UPF0004 | 1,044 | 1,044 | 0.95 | 0.97 | 0.96 |
| Ribosomal protein S12-S23 | 1,016 | 1,016 | 0.94 | 0.98 | 0.96 |
| GHMP kinases C terminal | 1,011 | 1,011 | 0.88 | 0.92 | 0.90 |
| Ribosomal protein S14p-S29e | 997 | 997 | 0.93 | 0.98 | 0.95 |
| Ribosomal protein S11 | 980 | 980 | 0.96 | 0.98 | 0.97 |
| UvrB-uvrC motif | 968 | 968 | 0.94 | 0.96 | 0.95 |
| Ribosomal protein L33 | 958 | 958 | 0.96 | 0.98 | 0.97 |
| BRCA1 C Terminus (BRCT) domain | 956 | 956 | 0.94 | 0.95 | 0.95 |
| RF-1 domain | 950 | 950 | 0.93 | 0.97 | 0.95 |
| Ankyrin repeats (3 copies) | 944 | 944 | 0.89 | 0.88 | 0.88 |
| Ribosomal protein L20 | 932 | 932 | 0.96 | 0.99 | 0.97 |
| RNA polymerase beta subunit | 912 | 912 | 0.94 | 0.97 | 0.95 |
| Ribosomal protein S18 | 908 | 908 | 0.93 | 0.97 | 0.95 |
| ATP synthase B-B CF(0) | 900 | 900 | 0.92 | 0.94 | 0.93 |
| Peptidase family M20-M25-M40 | 889 | 889 | 0.92 | 0.93 | 0.93 |
| Ribosomal protein L18e-L15 | 887 | 887 | 0.93 | 0.96 | 0.95 |
| Glucose inhibited division protein A | 886 | 886 | 0.95 | 0.96 | 0.95 |
| NADH-ubiquinone-plastoquinone oxidoreductase chain 4L | 885 | 885 | 0.94 | 0.97 | 0.96 |
| lactate-malate dehydrogenase, NAD binding domain | 880 | 880 | 0.92 | 0.94 | 0.93 |
| HD domain | 879 | 879 | 0.93 | 0.93 | 0.93 |
| Ribosomal protein S10p-S20e | 873 | 873 | 0.95 | 0.97 | 0.96 |

(Continued)

**Table 2.** (Continued)

| Family name | Training instances | | Classification Result | | |
|---|---|---|---|---|---|
| | # of positive sequences | # of negative sequences | Specificity | Sensitivity | Accuracy |
| Pyridoxal-phosphate dependent enzyme | 870 | 870 | 0.91 | 0.91 | 0.91 |
| Ribosomal L18p-L5e family | 860 | 860 | 0.93 | 0.96 | 0.94 |
| Ribosomal protein L3 | 855 | 855 | 0.94 | 0.97 | 0.96 |
| tRNA synthetases class I (M) | 843 | 843 | 0.94 | 0.96 | 0.95 |
| UbiA prenyltransferase family | 841 | 841 | 0.94 | 0.95 | 0.95 |
| Ribosomal protein L4–L1 family | 841 | 841 | 0.94 | 0.95 | 0.95 |
| Ribosomal protein S16 | 840 | 840 | 0.93 | 0.97 | 0.95 |
| Ribosomal protein S13-S18 | 840 | 840 | 0.94 | 0.97 | 0.95 |
| MraW methylase family | 837 | 837 | 0.95 | 0.98 | 0.96 |
| Ribosomal L32p protein family | 825 | 825 | 0.94 | 0.97 | 0.95 |
| Elongation factor TS | 819 | 819 | 0.94 | 0.97 | 0.96 |
| Tetrahydrofolate dehydrogenase-cyclohydrolase, catalytic domain | 817 | 817 | 0.94 | 0.96 | 0.95 |
| ATP synthase delta (OSCP) subunit | 813 | 813 | 0.93 | 0.96 | 0.94 |
| tRNA synthetases class I (C) catalytic domain | 812 | 812 | 0.95 | 0.97 | 0.96 |
| SecA Wing and Scaffold domain | 805 | 805 | 0.95 | 0.97 | 0.96 |
| Ribonuclease HII | 795 | 795 | 0.93 | 0.94 | 0.93 |
| Ribosomal protein L31 | 795 | 795 | 0.97 | 0.99 | 0.98 |
| Ribosomal L27 protein | 794 | 794 | 0.98 | 0.99 | 0.99 |
| IPP transferase | 794 | 794 | 0.93 | 0.95 | 0.94 |
| GTP-binding protein LepA C-terminus | 793 | 793 | 0.96 | 0.98 | 0.97 |
| Ribosomal protein L17 | 791 | 791 | 0.92 | 0.96 | 0.94 |
| Ribosomal protein L23 | 790 | 790 | 0.91 | 0.96 | 0.94 |
| Ribosomal protein L10 | 781 | 781 | 0.90 | 0.92 | 0.91 |
| Ribosomal protein L19 | 780 | 780 | 0.94 | 0.97 | 0.95 |
| Ribosomal protein S20 | 774 | 774 | 0.95 | 0.97 | 0.96 |
| Ribosomal protein L35 | 769 | 769 | 0.93 | 0.97 | 0.95 |
| Phosphoglucomutase-phosphomannomutase, C-terminal domain | 768 | 768 | 0.92 | 0.96 | 0.94 |
| AMP-binding enzyme | 767 | 767 | 0.87 | 0.89 | 0.88 |
| Ribosomal prokaryotic L21 protein | 766 | 766 | 0.93 | 0.96 | 0.95 |
| tRNA methyl transferase | 759 | 759 | 0.94 | 0.96 | 0.95 |
| Ribosomal L29 protein | 757 | 757 | 0.95 | 0.97 | 0.96 |
| Glycosyl transferase family, a-b domain | 754 | 754 | 0.90 | 0.91 | 0.91 |
| Translation initiation factor IF-2, N-terminal region | 750 | 750 | 0.96 | 0.98 | 0.97 |
| Ribosomal L28 family | 749 | 749 | 0.93 | 0.98 | 0.95 |
| Glycosyl transferase family 4 | 739 | 739 | 0.96 | 0.98 | 0.97 |
| tRNA synthetases class I (R) | 736 | 736 | 0.93 | 0.96 | 0.95 |
| Bacterial trigger factor protein (TF) C-terminus | 733 | 733 | 0.95 | 0.96 | 0.95 |
| For the first 1,000 families | 261,149 | 261,149 | 0.92 | 0.95 | 0.94 |
| For the first 2,000 families | 293,957 | 293,957 | 0.90 | 0.96 | 0.93 |
| For the first 3,000 families | 308,292 | 308,292 | 0.89 | 0.96 | 0.92 |
| For the first 4,000 families | 316,135 | 316,135 | 0.87 | 0.96 | 0.91 |
| Weighted average for all 7,027 families | 324,018 | 324,018 | 0.91 | 0.95 | 0.93 |

doi:10.1371/journal.pone.0141287.t002

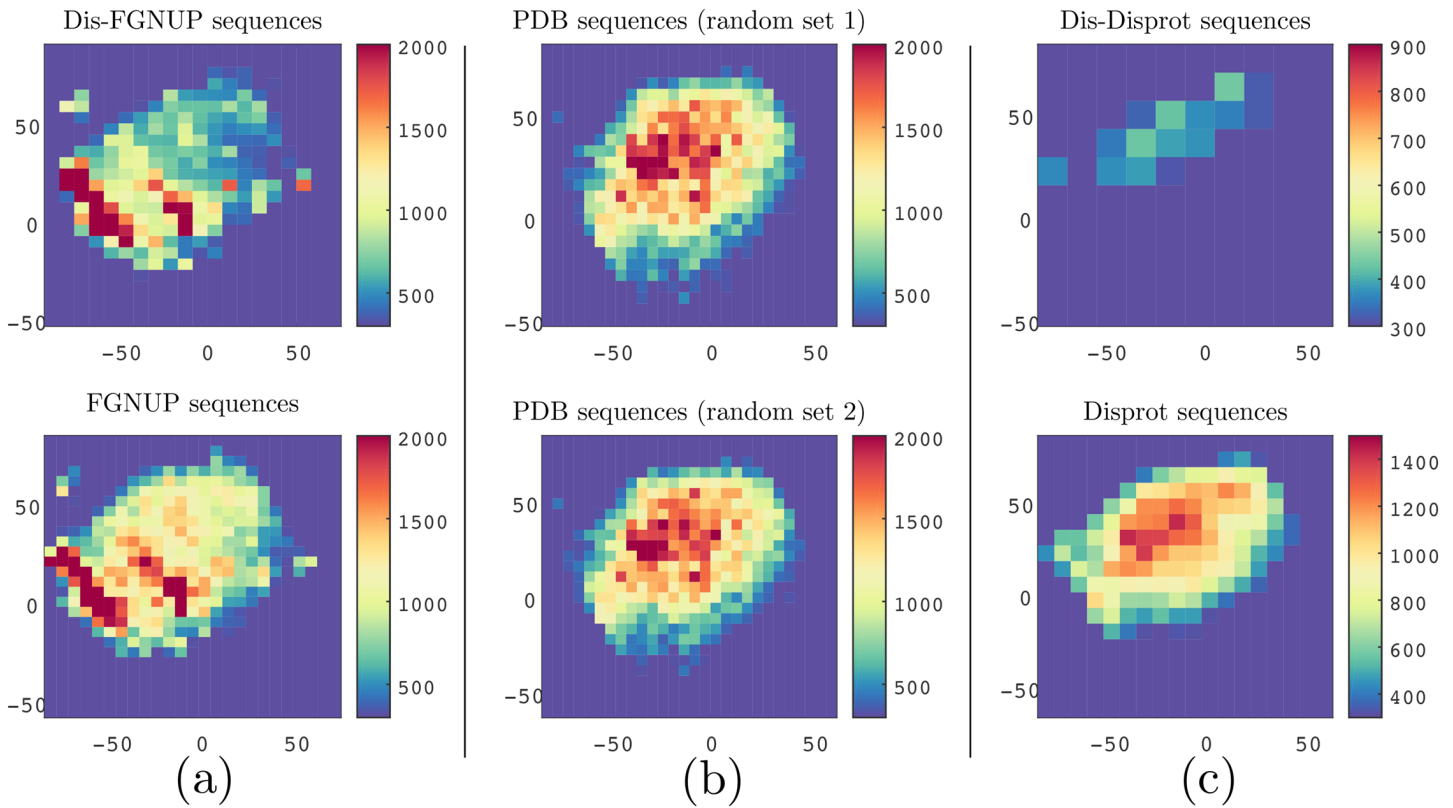

**Fig 3. Visualization of protein sequences using ProtVec can characterize FGNUPs versus Disport disordered sequences and structured sequences.** Column (a) compares FG Nup sequences 2D histogram (at the bottom) with 2D histogram of FG Nup disordered regions (on top). Column (b) compares 2D histogram two random sets of structured sequences with the same average length as the FG-Nups. Column (c) compares between 2D histogram of DisProt sequences (at the bottom) and 2D histogram of DisProt disordered regions (on top).

doi:10.1371/journal.pone.0141287.g003

In order to visualize each dataset, we reduced the dimensionality of the protein-space using Stochastic Neighbor Embedding [36, 39] and then generated the 2D histogram of all overlapping 3-grams occurring in each dataset. As shown in Fig 3 (see column (b)), the two random sets from structured proteins had nearly identical patterns. However, the FG-Nups dataset exhibits a substantially different pattern. To amplify the characteristic of disordered sequences we have also examined the histogram of disordered regions of FG-Nups (see Fig 3, column (a)).

In the next step, we quantitatively evaluated how ProtVec can be used to distinguish between FG-Nups versus typical PDB sequences using a support vector machine binary classification. The positive examples were the above mentioned 1,138 disordered FG-Nups proteins and negative examples (again 1,138 sequences) were selected randomly from PDB with the same average length of disordered sequences ($\approx 900$ residues). We represented each protein sequence as a summation of its ProtVecs of all 3-grams. Since on average the length of structured proteins were shorter than FG-Nups, in order to avoid trivial cases, the PDB sequences were selected in such a way as to maintain the same average length. But still, an accuracy of 99.81% was obtained with high sensitivity and specificity (Table 3). The distribution of the classified proteins in a 2D space is shown in Fig 4.

**Table 3. The performance of FG-Nups disordered protein classification in a 10xFold cross-validation using SVM.**

| Sensitivity | Specificity | Accuracy |
|---|---|---|
| 0.9987 | 0.9974 | 0.9981 |

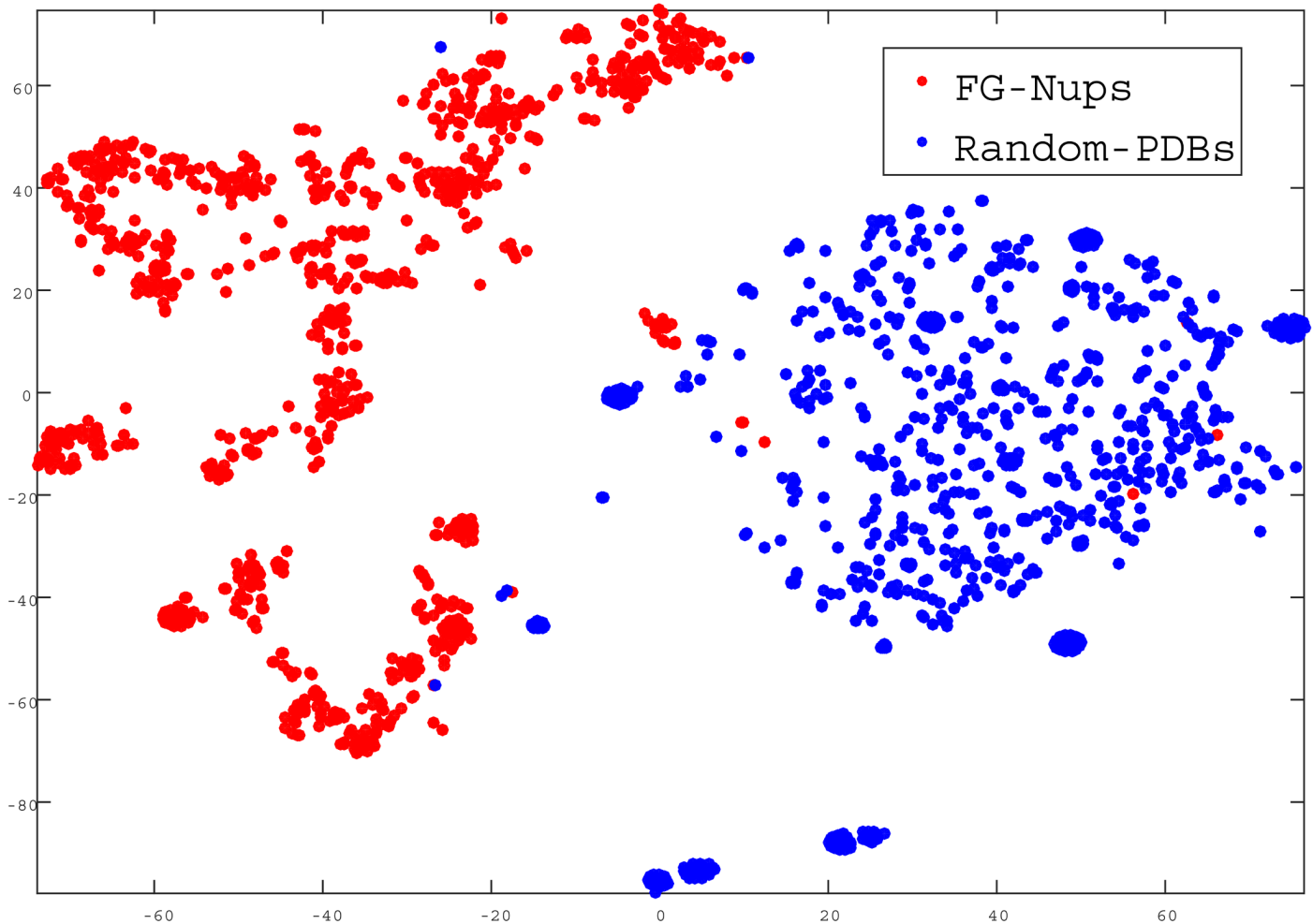

**Fig 4. Classification of FG-Nups versus PDB structured sequences.** In this figure, each point presents a protein projected into a 2D space.

**DisProt characterization.** In this part, we used DisProt consisting of 694 proteins presenting 1539 disordered, and 95 ordered regions. We performed the same analysis as we did for FG-Nups with DisProt sequences (see Fig 3 column (c)). Since the size of DisProt was relatively small compared to that of the FG-Nups, the scales of columns (a),(b) were not comparable with column (c) (see Fig 3). The visualization of disordered regions of DisProt sequences (Fig 3 column (a), on top) revealed a different characteristic than FG-Nups disordered regions (Fig 3 column (a), on top). A visual comparison between Figs 3 and 2 suggest that the FG-Nups have a significantly higher amount of hydrophobic residues and less polar residues in their disordered regions than the experimentally identified disordered proteins in DisProt [27, 29]. Additionally, the DisProt disordered regions versus DisProt ordered regions can be classified with 100% accuracy respectively using SVM and ProtVec.

## Conclusions

An unsupervised data-driven distributed representation, called ProtVec, was proposed for application of machine learning approaches in biological sequences. By training this representation solely on protein sequences, our feature extraction approach was able to capture a diverse range of meaningful physical and chemical properties. We demonstrated that ProtVec can be used as an informative and dense representation for biological sequences in protein family classification, and obtained an average family classification accuracy of 93%.

We further proposed ProtVec as a powerful approach for protein data visualization and showed the utility of this approach by providing an example in characterization of disordered protein sequences vs. structured protein sequences. Our results suggest that ProtVec can characterize protein sequences in terms of biochemical and biophysical interpretations of the underlying patterns. In addition, this dense representation of sequences can help to discriminate between various categories of sequences, e.g. disordered proteins. Furthermore, we demonstrated that ProtVec was able to identify disordered sequences with an accuracy of nearly 100%. The related data is available at: http://llp.berkeley.edu and Harvard Dataverse: http://dx.doi.org/10.7910/DVN/JMFHTN.

Another advantage of this method is that embeddings could be trained once and then used to encode biological sequences in any given problem. In general, machine learning approaches in bioinformatics can widely benefit from bio-vectors (ProtVec and GeneVec) representation. This representation can be considered as pre-training for various applications of deep learning in bioinformatics. In particular, ProtVec can be used in protein interaction predictions, structure prediction, and protein data visualization.

## Supporting Information

**S1 File. The results of family classification task for all 7,027 families.**
(XLSX)

## Acknowledgments

Fruitful discussions with Kiavash Garakani, Mohammad Soheilypour, Zeinab Jahed, Mohaddeseh Peyro, Hengameh Shams and other members of the Molecular Cell Biomechanics Lab at the University of California Berkeley are gratefully acknowledged.

## Author Contributions

Conceived and designed the experiments: EA MRKM. Performed the experiments: EA. Analyzed the data: EA MRKM. Contributed reagents/materials/analysis tools: MRKM. Wrote the paper: EA MRKM.